\documentclass[pra, twocolumn, showpacs, superscriptaddress]{revtex4}

\usepackage{graphicx,amsmath,amssymb}
\usepackage{gensymb}

\begin{document}

\title{Collisionless spin dynamics in a magnetic field gradient }
\author{Junjun Xu}
\affiliation{Department of Physics, University of Science and Technology Beijing, Beijing 100083, China}
\affiliation{Laboratory of Atomic and Solid State Physics, Cornell University, Ithaca, New York 14853, USA}
\author{Qiang Gu}\email{qgu@ustb.edu.cn}
\affiliation{Department of Physics, University of Science and Technology Beijing, Beijing 100083, China}
\author{Erich J. Mueller}\email{em256@cornell.edu}
\affiliation{Laboratory of Atomic and Solid State Physics, Cornell University, Ithaca, New York 14853, USA}
\date{\today}

\begin{abstract}
We study the collisionless spin dynamics of a harmonically trapped Fermi gas in a magnetic field gradient. In the absence of interactions, the system evolution is periodic: the magnetization develops twists, which evolve into a longitudinal polarization.  Recurrences follow.  For weak interaction, the exchange interactions lead to beats in these oscillations. We present an array of analytic and numerical techniques for studying this physics.
\end{abstract}

\maketitle

\section{Introduction}
The spins in Fermi gases display a rich range of behaviors: Both collisionless and collision dominated spin waves have been studied in a number of recent experiments \cite{Sommer, Sommer2, Wulin, Koschorreck, Krauser, Bardon, Hild}. Here we present a simple model of the collisionless regime, and  study the dynamics.

Collisionless spin waves were first explored in $^3$He film, then in spin polarized hydrogen \cite{Lhuillier, Bigelow, Schafer}. However, some of these early works are hard to observe in these atomic systems. Unlike these early studies, cold atom experiments allow one to directly image the spin waves \cite{Bloch, Ketterle}. Such images complement the spectroscopic and transport probes used to understand $^3$He and hydrogen.

While our focus is the weakly interacting regime, much recent work has explored the collision dominated strongly interacting limit. One of the problems is to understand the spin diffusive behavior in this regime. For example, it is conjectured that quantum mechanics forbids the spin diffusion constant from exceeding $\sim\hbar/m$, where $\hbar$ is the reduced Planck's constant and $m$ is the atomic mass. This bound appears to be consistent with the recent theoretical and experiment results\cite{Sommer, Sommer2, Wulin,  Koschorreck, Bardon, Hild, Enss1, Enss2, Bruun1, Bruun2, Wlazlowski}.

Our model is largely inspired by the experiments of Bardon {\it et al}. in Toronto \cite{Bardon}. There a two-component Fermi gas is placed in a cigar shape trap. The gas is prepared with a uniform magnetization in the $\hat x$ direction.  A magnetic field is applied in the $\hat z$ direction.  The strength of this field varies linearly with $z$.  The experimentalists let the system evolve for some time then interrogate it, typically with a spin-echo protocol \cite{Hahn, Abragam}. For simplicity, we do not model the spin-echo, but simply study how the magnetization evolves. Since all dynamics are one-dimensional (1D), and the Hamiltonian can be integrated out in the other two dimensions, we will focus on the 1D model instead.

This paper is organized as follows. First we consider the non-interacting case, where we can analytically describe the dynamics. We show that the transverse magnetization oscillates, and explain this behavior in terms of transverse and longitudinal spin rotations. Then we study weak interaction. We find beats which can be attributed to the exchange interaction. In the end, we discuss the transition from collisionless to the collisional limit and give our summary.

\section{Non-interacting case}
We consider 1D pseudospin-1/2 Fermi gases in a  magnetic field gradient along the $x$ axis.  In the absence of interactions, we can consider the dimensionless single-particle Hamiltonian
\begin{eqnarray}
H_0=-\frac{\partial_x^2}{2}+\frac{x^2}{2}+\sigma\lambda x,
\label{eq:hamiltonian}
\end{eqnarray}
where $\sigma=\pm$ for the up and down spin states, with respect to the $\hat z$ axis. Here $x=\tilde{x}/\sqrt{\hbar/(m\omega)}$, where $\tilde{x}$ is the position of the particle and $\sqrt{\hbar/(m\omega)}$ is the characteristic length of the harmonic oscillator, with $\omega$ the trap frequency.  The magnetic field gradient is represented by dimensionless $\lambda$. The actual magnetic gradient is $\tilde{\lambda}=\lambda\omega\sqrt{\hbar m\omega}/\mu$, where $\mu$ is the magnetic moment.  The dimensionless time $t=\omega\tilde{t}$, where $\tilde{t}$ is the real time. Hereafter we will use these dimensionless quantities. The many-body state will consist of a Slater determinant of single-particle states $\psi_{n\sigma}(x,t)$, each of which evolve via the time-dependent Schrodinger equation, $i \partial_t \psi_n=H \psi_n$.  The index $n$ runs from $1$ to $N$, where $N$ is the number of particles.

We envision preparing the system so that at time $t=0$ the $n$'th state of the Slater determinant has spin components $\psi_{n\uparrow}= \phi_n(x) /\sqrt{2}$ and $\psi_{n\downarrow}= \phi_n(x) /\sqrt{2}$, where $\phi_n(x)$ is the $n$'th eigenstate of the simple harmonic oscillator.  Such an initial condition is prepared by first polarizing the system in the $\uparrow$ state, and cooling it to the ground state.  One then applies a $\pi/2$ pulse, rotating the spin from the $\hat z$ direction to the $\hat x$ direction.

To find the subsequent dynamics, we make the following ansatz:
\begin{eqnarray}
\psi_{n\sigma}\left(x,t\right)=\phi_n\left(x-\sigma x_0\left(t\right)\right)e^{i\sigma xv\left(t\right)}e^{i\Phi_n\left(t\right)}.
\end{eqnarray}
This ansatz describes the exact dynamics in the absence of interactions. Substituting this ansatz into the Schrodinger equation yields a set of ordinary differential equations for $x_0$, $v$ and $\Phi_n$,
\begin{eqnarray}
&&x_0''(t)+x_0(t)+\lambda=0,\\
&&x_0^2(t)/2-v^2(t)/2-\Phi_n'(t)-E_n=0.
\end{eqnarray}
These equations are readily integrated to yield
\begin{eqnarray}
x_0\left(t\right)&=&\lambda\left(\cos t-1\right), v\left(t\right)=-\lambda \sin t,\\
\Phi_n\left(t\right)&=&\frac{\lambda^2 \sin2t}{4}-\lambda^2\sin t-\left(n+\frac{1}{2}-\frac{\lambda^2}{2}\right)t.
\end{eqnarray}
The local magnetization in the $x,y$ plane can be expressed as a complex number $m(x,t) = m_x+i m_y$.  This complex magnetization can be written as
$m(x,t)=\sum_n m_n(x,t)$ with
\begin{eqnarray}
m_n\left(x,t\right)&=&\psi_{n\downarrow}^*\left(x,t\right)\psi_{n\uparrow}\left(x,t\right)\nonumber\\
&=&\phi_n\left(x-x_0\left(t\right)\right)\phi_n\left(x+x_0\left(t\right)\right)e^{2ixv\left(t\right)}.\label{m2}
\end{eqnarray}

\begin{figure}[h]
  \includegraphics[width=.5\textwidth]{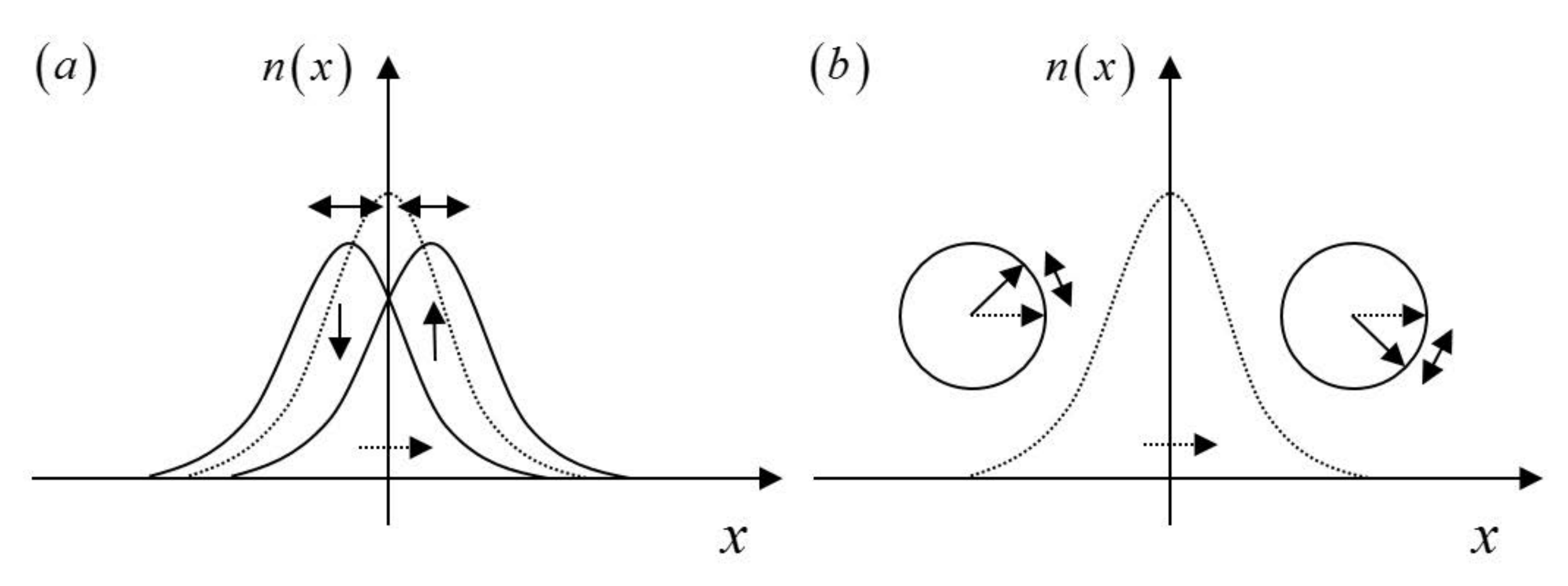}\\
  \caption{Illustration of the processes which lead to magnetization dynamics of a Fermi gas in a magnetic field gradient.  (a) Longitudinal spin dynamics: The up-spin and down-spin atoms move relative to each other, changing their overlap, and hence the transverse magnetization.  (b) Transverse spin dynamics: The spins precess in the transverse plane, at speeds which depend on the location of the atoms.  This leads to an inhomogeneous texture, whose average magnetization is reduced. The single arrow labels the spin direction, while the double one represents the movement of the spin.}\label{fig:fig1}
\end{figure}

We give the physical picture leading to magnetization dynamics in this non-interacting case in Fig. \ref{fig:fig1}. As illustrated in Fig.~\ref{fig:fig1}(a), the center of the up-spin and down-spin clouds become separated by a distance $2 x_0(t)$.  This reduces the overlap between the clouds, and the magnitude of the local transverse magnetization.  As is evident in Eq.~(\ref{m2}), the clouds also develop a relative phase profile $\exp(2ixv)$.  As shown in Fig.~\ref{fig:fig1}(b), this phase factor can be interpreted as a spin precession term.  While it does not change the magnitude of the local polarization, it makes the direction depend on position.  Thus it does reduce the total transverse magnetization $M(t)=\int m(x,t)\,dx$.  By symmetry $M(t)$ is always real, implying the net transverse polarization is always in the $\hat x$ direction.  By construction $M(t)\leq N$ is dimensionless.

\begin{figure}[htb]
  \includegraphics[width=.4\textwidth]{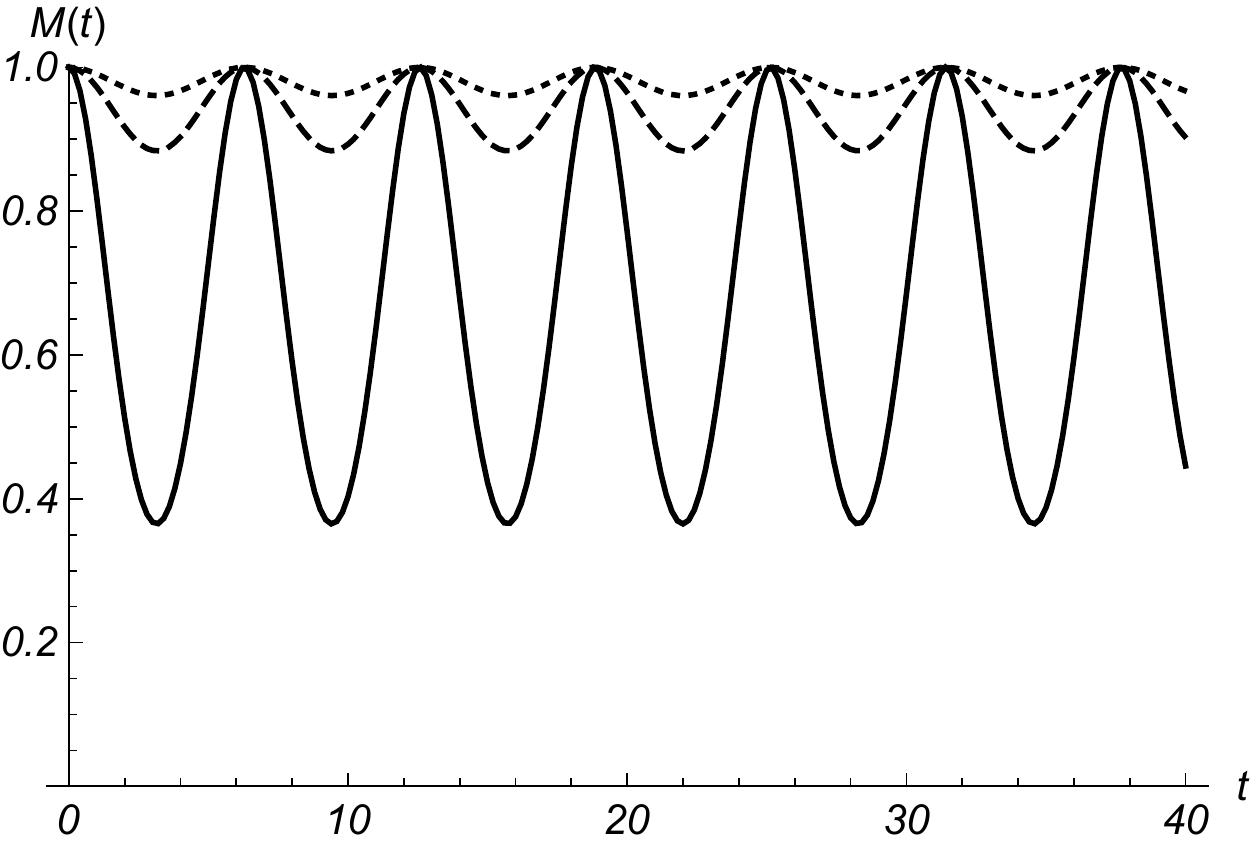}\\
  \caption{Time evolution of average transverse magnetization per particle $M(t)/N$ (solid) for $N=21$ particles. Here $t=\omega\tilde{t}$ is the reduced time, where $\tilde{t}$ is the real time and $\omega$ is the trap frequency. Also shown is the average magnetization of the particles for the two lowest states: $n=0$ (dotted) and $n=1$ (dashed).}\label{fig:fig2}
\end{figure}

Figure~\ref{fig:fig2} shows the evolution of the transverse magnetization for $N=21$ particles, taking $\lambda=0.1$.  As expected from Eq.~(\ref{m2}), we see clear undamped oscillations.  We also show the contribution from the $n=0$ and $n=1$ states.  The lower energy states dominate the dynamics, as the maximum displacement $|x_0(t=\pi)|=2\lambda$ is a greater fraction of their width.

\section{weak interaction}
In this section, we consider the influence of weak interaction on the system. We make a time-dependent Hartree-Fock ansatz, assuming that at all times the system is described by a Slater determinant of $N$ single-particle states.  The equations of motion can be derived by minimizing the action
\begin{equation}
S=\int \Psi^*(i\partial_t-H)\Psi\,dt\, d^Nx,
\end{equation}
where $\Psi(x_1,x_2,\cdots,x_N,t)$ represents the many-body wave function.  The action can be decomposed as a sum of a non-interacting and an interacting part
$S=S_0+S_I$.  Within the Hartree-Fock ansatz,
\begin{eqnarray}
S_0
&=&\sum_{n,\sigma}\int \psi_{n,\sigma}^*\left(i\partial_t+\frac{\partial_x^2}{2}-\frac{x^2}{2}-\sigma\lambda x\right)\psi_{n,\sigma}dxdt,\nonumber\\
S_I&=&\sum_{m,n}\int g\left(\vert\psi_{m,\uparrow}\vert^2\vert\psi_{n,\downarrow}\vert^2-\psi_{m,\uparrow}^*\psi_{m,\downarrow}\psi_{n,\downarrow}^*\psi_{n,\uparrow}\right)dxdt\nonumber
\end{eqnarray}
with $g=\tilde{g}/(\hbar\omega\sqrt{\hbar/(m\omega)})$ represents the reduced interaction strength with $\tilde{g}$ the physical one. Minimizing the action with respect to the wave functions $\psi_{n,\sigma}^*(x)$ yields equations of motion
\begin{eqnarray}
i\partial_t\psi_{n,\sigma}&=&H_0\psi_{n,\sigma}+g\sum_m\big(\vert\psi_{m,\bar{\sigma}}\vert^2\psi_{n,\sigma}\nonumber\\
&-&\psi_{m,\sigma}^*\psi_{m,\bar{\sigma}}\psi_{n,\bar {\sigma}}\big).
\label{eq:hf}
\end{eqnarray}
As argued in \cite{Baym}, in the limit of slow spatial and temporal dynamics, these coupled equations are equivalent to the collisionless Boltzmann equation.
For moderate $N$, we can numerically integrate these equations.  The exchange interaction acts like a spatially inhomogeneous transverse magnetic field, and scrambles the spins.  The simple oscillations seen in Fig.~\ref{fig:fig2} develop more structure, exhibiting quantum beats as seen in Fig.~\ref{fig:fig4}.

These beats can be qualitatively understood by considering a single wave function $\psi_{n,\sigma}(x)$ in the Slater determinant, and treating the other wavefunctions as a static homogeneous background.  Within this approximation, Eq.~(\ref{eq:hf}) becomes a similar Jaynes-Cummings model,
\begin{eqnarray}\label{eq:approx}
i\partial_t\left(\begin{array}{c}\psi_{n,\uparrow}\\ \psi_{n,\downarrow} \end{array}\right)=\bar{H}\left(\begin{array}{c}\psi_{n,\uparrow}\\ \psi_{n,\downarrow} \end{array}\right),
\end{eqnarray}
where
\begin{eqnarray}
\bar{H}=-\frac{\partial^2_x}{2}+\frac{x^2}{2}+\lambda x\sigma_z+g_e\sigma_x,
\end{eqnarray}
with $g_e$ the effective exchange interaction strength. The subsequent mathematics is simpler if we rotate spin space by $90$\degree. This is done by introducing a transformation matrix
\begin{equation}
R=\frac{1}{\sqrt{2}}\left(
\begin{array}{cc}
1 & 1 \\
1 & -1 \\
\end{array}
\right),
\end{equation}
and defining
\begin{eqnarray}
\left(\begin{array}{c}\psi_{n,\rightarrow}\\\psi_{n\leftarrow}\end{array}\right)&=&
R \left(\begin{array}{c}\psi_{n,\uparrow}\\
\psi_{n\downarrow}\end{array}\right),\\
\bar H^\prime&=&R \bar{H} R.
\end{eqnarray}
 Equation (\ref{eq:approx}) then becomes
\begin{eqnarray}\label{eq:simmodel}
i\partial_t\left(\begin{array}{c}\psi_{n,\rightarrow}\\\psi_{n,\leftarrow}\end{array}\right)=\bar{H}^\prime
\left(\begin{array}{c}\psi_{n,\rightarrow}\\\psi_{n,\leftarrow} \end{array}\right),
\end{eqnarray}
where
\begin{eqnarray}
\bar{H}^\prime=\left(\begin{array}{cc}
(n+\frac{1}{2})+g_e&\frac{\lambda}{\sqrt{2}}(a^\dagger+a)\\
\frac{\lambda}{\sqrt{2}}(a^\dagger+a)&(n+\frac{1}{2})-g_e
\end{array}\right),
\end{eqnarray}
with $a=(x+\partial_x)/\sqrt{2}, a^\dagger=(x-\partial_x)/\sqrt{2}$. In this rotated vector space our initial state is $\psi_{n,\rightarrow}(t=0)= \phi_n(x)$, $\psi_{n\leftarrow}(t=0)=0$.

\begin{figure}[t]
  \includegraphics[width=.40\textwidth]{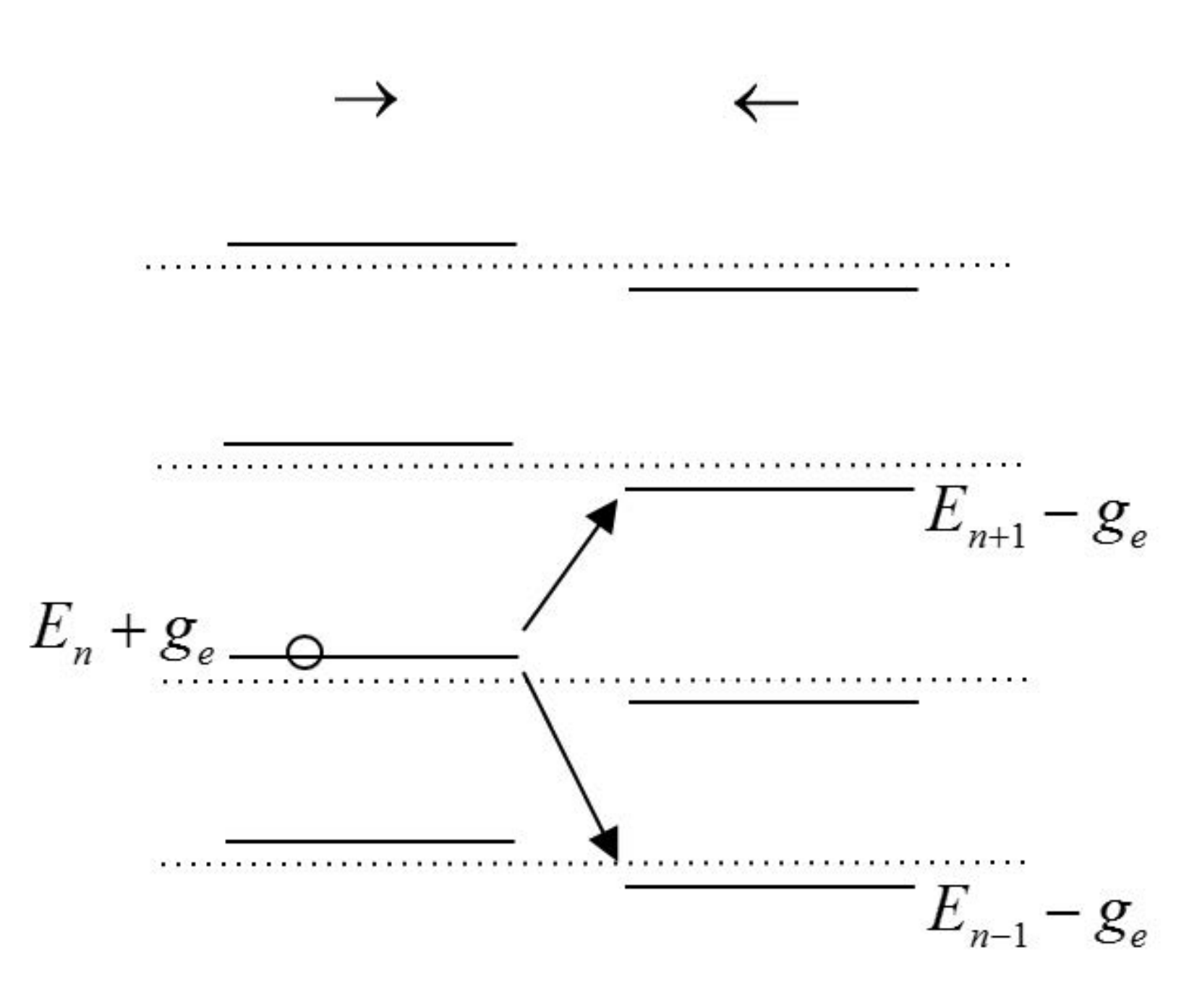}\\
  \caption{Level structure of the simplified model in Eq. (\ref{eq:simmodel}), used to describe the time evolution of a single wavefunction $\psi_{n,\sigma}(x)$ in the Slater determinant. Dashed lines show the eigen-energies when the field gradient $\lambda$ and effective exchange field $g_e$ are both zero.  The exchange field splits the degeneracy between $\rightarrow$ and $\leftarrow$ states, yielding the solid lines. $E_n=n+1/2$. The gradient $\lambda$ couples levels as shown by the arrows.  At time $t=0$, $\psi_{n,\rightarrow}(x)=\phi_n(x)$ and $\psi_{n,\leftarrow(x)}=0$, corresponding to the state marked with a circle.}\label{fig:fig3}
\end{figure}

This Hamiltonian  can be understood  from the level diagram in Fig. \ref{fig:fig3}. For zero interaction and magnetic field gradient, the system is a simply harmonic oscillator and has energy levels at $E_n=n+1/2$, which is the dashed lines in Fig. \ref{fig:fig3}. The exchange coupling $g_e$ shifts the eigenstate energy of $\rightarrow$ and $\leftarrow$ spins, shown as the solid lines and energy $E_n\pm g_e$. The magnetic field gradient flips the spins and at the same time changes the vibrational level by one.  For small $\lambda$, one can truncate the time dynamics to a three level system: the initial state, labeled by a circle, and the two states which are coupled to it.  This yields two oscillation frequencies $\nu_{\pm}=1\pm2 g_e$, and to second order in $\lambda$ the contribution to the magnetization from the $n$'th term in the Slater determinant is
$m_n=1+\frac{(n+1)\lambda^2}{(1-2g_e)^2}\cos(1-2g_e)t+\frac{n\lambda^2}{(1+2g_e)^2}\cos(1+2g_e)t$.  The beats in Fig.~\ref{fig:fig4}  are qualitatively consistent with this scenario, but this simplified model is not able to quantitatively describe the evolution of the magnetization. A more sophisticated, and quantitatively accurate, ansatz involves taking the $n$'th wavefunction in the Slater determinant as having the form
\begin{eqnarray}
\psi_{n,\sigma}(x,t)&=&A_n(t)\phi_n(x)\nonumber\\
 &+&\sigma B_n(t)\phi_{n+1}(x)+\sigma C_n(t)\phi_{n-1}(x).
\end{eqnarray}
Minimizing the action with this ansatz yields equations of motion for $A_n,B_n,$ and $C_n$,
\begin{eqnarray*}
\left[\left(\begin{array}{ccc}
i\partial_t - E_n&-\lambda X_n&-\lambda X_{n-1}\\
-\lambda X_n&i\partial_t-E_{n+1}&\\
-\lambda X_{n-1}&&i\partial_t-E_{n-1}
\end{array}\right)\right.\\
\left.+4g\sum_m\Lambda_{mn}\right]
\left(\begin{array}{c}
A_n\\B_n\\C_n
\end{array}\right)=0,
\end{eqnarray*}
where $X_n=\int dx\phi_n(x)\phi_{n+1}(x)$. The nonlinear term is
\begin{widetext}
\begin{equation}
\Lambda_{mn}=
\left(\begin{array}{ccc}
\vert B_m\vert^2\alpha^{m+1}_n+\vert C_m\vert^2\alpha^{m-1}_n+2Re(B_m^*C_m)\beta^n_m&-A_m(B_m^*\gamma^m_n+C_m^*\gamma^{m-1}_n)&-A_m(B_m^*\gamma^m_{n-1}+C_m^*\gamma^{m-1}_{n-1})\\
-A_m^*(B_m\gamma^m_n+C_m\gamma^{m-1}_n)&\vert A_m\vert^2\alpha^m_{n+1}&\vert A_m\vert^2\beta^m_n\\
-A_m^*(B_m\gamma^m_{n-1}+C_m\gamma^{m-1}_{n-1})&\vert A_m\vert^2\beta^m_n&\vert A_m\vert^2\alpha^m_{n-1}
\end{array}\right)
\end{equation}
\end{widetext}
where
\begin{eqnarray*}
\alpha^m_n&=&\int dx\phi_m^2(x)\phi_n^2(x),\\
\beta^m_n&=&\int dx\phi_m^2(x)\phi_{n+1}(x)\phi_{n-1}(x),\\
\gamma^m_n&=&\int dx\phi_m(x)\phi_{m+1}(x)\phi_n(x)\phi_{n+1}(x).
\end{eqnarray*}
The above equations can be solved iteratively and we get the net magnetization
\begin{eqnarray}
M(t)=\sum_n\left(\vert A_n(t)\vert^2-\vert B_n(t)\vert^2-\vert C_n(t)\vert^2\right).
\end{eqnarray}

We plot the magnetization for system $N=21$ and $\lambda=0.1$ in Fig. \ref{fig:fig4}. The reduced interaction is chosen as $g=0.1$. We can see the weak interaction develops beats in the magnetization, which is expected from our previous approximation. These differential equations are much easier to integrate than the Hartree-Fock equations, and
as seen in Fig. \ref{fig:fig4}, this simple ansatz captures most of the relevant physics.
\begin{figure}[tb]
  \includegraphics[width=.4\textwidth]{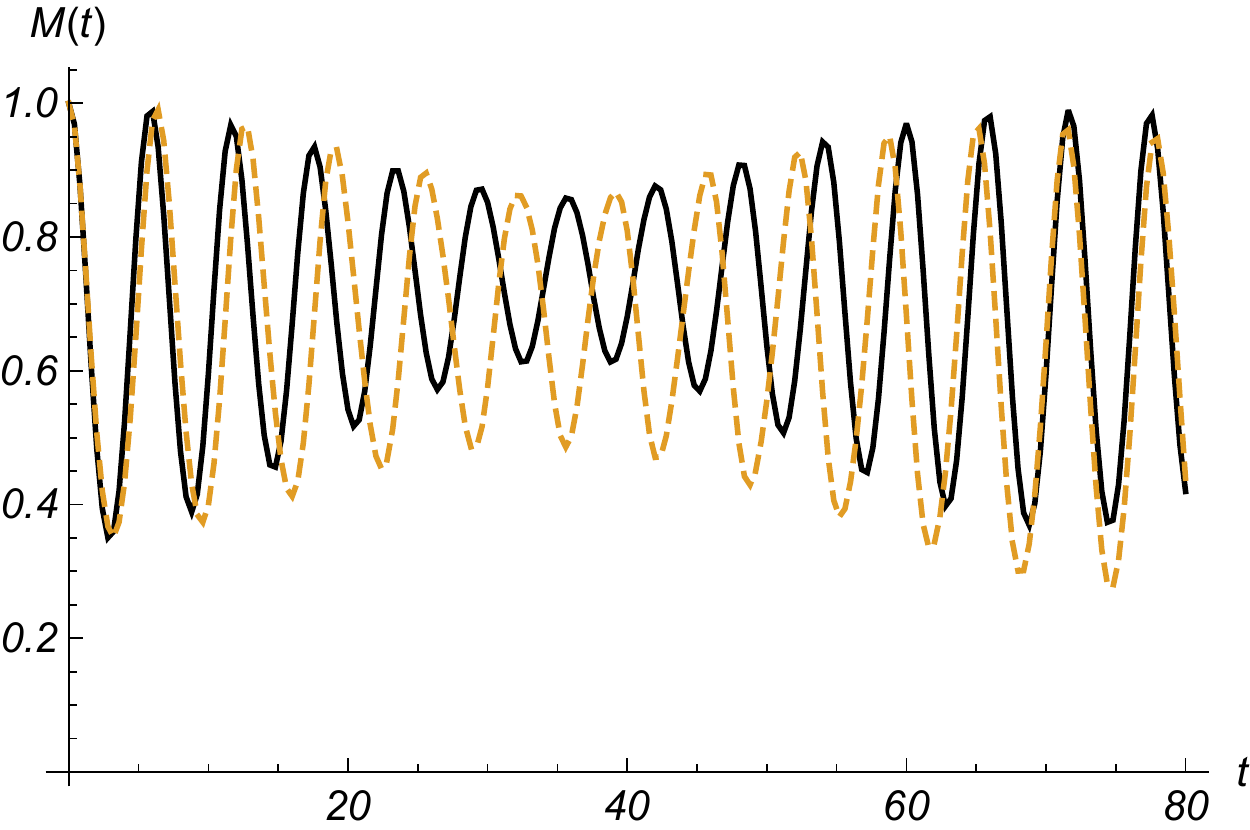}\\
  \caption{(Color online) Time evolution of average transverse magnetization for reduced interaction strength $g=\tilde{g}/(\hbar\omega\sqrt{\hbar/(m\omega)})=0.1$. The particle number and magnetic gradient is the same as the non-interacting case and $t=\omega\tilde{t}$ is the reduced time with $\tilde{t}$  the real time and $\omega$ the trap frequency. Solid black line: Simplified variational approximation.   Dashed orange line: Full Hartree-Fock calculation.}\label{fig:fig4}
\end{figure}

Finally, we give the condition when the actual three-dimensional (3D) experiment goes from collisionless to collisional limit. As argued in \cite{Jin}, this transition is determined by $\Gamma/\omega$, where $\Gamma=n\sigma v$ is the two-body elastic scattering rate per particle with $n$ the mean density of the system, $\sigma$ the scattering cross section, and $v$ the relative velocity. If the scattering event seldom happens at several oscillation periods, i.e.,  $\Gamma\ll\omega$, the dynamics is mainly dominated by collisionless processes. The system shows spin wave dynamics in this regime. Increasing the scattering rate $\Gamma$, it continuously goes from collisionless limit to collisional limit. In our system $n\approx (2mE_F)^{3/2}/(6\pi^2\hbar^3)\approx43.8(m\omega/\hbar)^{3/2}/\pi^2$ for $E_F=20.5\hbar\omega$, and $v\sim\lambda\sqrt{\hbar\omega/m}$. Our 1D reduced interaction strength is connected to the 3D scattering length as $\tilde{g}=g\hbar\omega\sqrt{\hbar/(m\omega)}\approx2\hbar^2a_{3D}/(ml^2)$ \cite{Guan}. Here $l=\sqrt{\hbar/(m\omega_\perp)}$ is the transverse width and for a isotropic harmonic potential we have $a_{3D}=g\sqrt{\hbar/(m\omega)}/2$. Thus we have $\Gamma=4\pi na_{3D}^2v\sim 43.8g^2\lambda\omega/\pi$. So for a strongly interacting collisional limit, one has to fulfill $g\gg\sqrt{\pi/(43.8\lambda)}\approx 0.8$ for reduced magnetic gradient $\lambda=0.1$. Beyond this point, the system is dominated by collisional processes and the physics of spin diffusion begin to take part.

\section{summary}
In summary, we give an analysis of the collisionless spin dynamics of a weakly interacting Fermi gas in a magnetic field gradient.  In the absence of interactions the net transverse magnetization oscillates with the trap frequency.  Weak interactions lead to beats.  We present simple models which explain these dynamics.

For a related experiment, the magnetization dynamics can be directly observed using the techniques in Ref. \cite{Bardon}, and the beats behavior clearly tells the low energy spin wave dynamics of the system. Here even though our calculation is purely 1D, we believe the beats behavior is also preserved in the 3D system. This can also be seen from our mean-field argument, where the other dimensions will only change the eigenenergy of the harmonic oscillator. However the other two dimensions do contribute to the system. They will influence the local interaction strength and then shift the beats. Also in the experiment one needs to take into account the effect of temperature. For a non-zero temperature, the atoms will spread to higher single-particle states and the spin dynamics will be enhanced.

\begin{acknowledgements}
This research is supported by the National Key Basic Research Program of China (Grant No. 2013CB922002), the National Natural Science Foundation of China (Grant No. 11074021), and the ARO-MURI Non-equilibrium Many-body Dynamics Grant (W911NF-14-1-0003). J.X. is also supported by China Scholarship Council.
\end{acknowledgements}


\begin{thebibliography}{50}

\bibitem{Sommer} A. Sommer, M. Ku, G. Roati, and M. W. Zwierlein, Nature {\bf 472}, 201 (2011).

\bibitem{Sommer2} A. Sommer, M. Ku, and M. W. Zwierlein, New J. Phys. {\bf 13}, 055009 (2011).

\bibitem{Wulin} D. Wulin, H. Guo, C. C. Chien, and K. Levin, Phys. Rev. A {\bf 83}, 061601 (2011).

\bibitem{Koschorreck} M. Koschorreck, D. Pertot, E. Vogt, and M. K\"{o}hl, Nat. Phys. {\bf 9}, 405 (2013).

\bibitem{Bardon} A. B. Bardon, S. Beattie, C. Luciuk, W. Cairncross, D. Fine, N. S. Cheng, G. J. A. Edge, E. Taylor, S. Zhang, S. Trotzky, and J. H. Thywissen, Science {\bf 344}, 722 (2014).

\bibitem{Hild} S. Hild, T. Fukuhara, P. Schau{\ss}, J. Zeiher, M. Knap, E. Demler, I. Bloch, and C. Gross, Phys. Rev. Lett. {\bf 113}, 147205 (2014).

\bibitem{Krauser} J. S. Krauser, U. Ebling, N. Fl\"{a}schner, J. Heinze, K. Sengstock, M. Lewenstein, A. Eckardt, and C. Becker, Science {\bf 343}, 157 (2014).

\bibitem{Lhuillier} C. Lhuillier et F. Lalo\"{e}, J. Phys. France {\bf 43}, 197 (1982).

\bibitem{Bigelow} N. P. Bigelow, J. H. Freed, and D. M. Lee, Phys. Rev. Lett. {\bf 63}, 1609 (1989).

\bibitem{Schafer} T. Sch\"{a}fer and D. Teaner, Rep. Prog. Phys. {\bf 72}, 126001 (2009).

\bibitem{Bloch} I. Bloch, J. Dalibard, and W. Zwerger, Rev. Mod. Phys. {\bf 80}, 885 (2008).

\bibitem{Ketterle} W. Ketterle and M. W. Zwierlein, {\it Ultracold Fermi Gases}, Proceedings of the International School of Physics "Enrico Fermi", Course CLXIV, Varenna, 2006, edited by M. Inguscio, W. Ketterle, and C. Salomon (IOS Press, Amsterdam, 2008).

\bibitem{Enss1} T. Enss, R. Haussmann, Phys. Rev. Lett. {\bf 109}, 195303 (2012).

\bibitem{Enss2} T. Enss, Phys. Rev. A {\bf 88}, 033630 (2013).

\bibitem{Bruun1} G. M. Bruun, New. J. Phys. {\bf 13}, 035005 (2011).

\bibitem{Bruun2} G. M. Bruun, Phys. Rev. A {\bf 85}, 013636 (2012).

\bibitem{Wlazlowski} G. Wlazlowski, P. Magierski, J. E. Drut, A. Bulgac, and K. J. Roche, Phys. Rev. Lett. {\bf 110}, 090401 (2013).

\bibitem{Hahn} E. L. Hahn, Phys. Rev. {\bf 80}, 580 (1950).

\bibitem{Abragam} A. Abragam, \textit{The Principles of Nuclear Magnetism} (Oxford University Press, London, 1961).

\bibitem{Baym} L. P. Kadanoff and G. Baym, \textit{Quantum Statistical Mechanics} (W. A. Benjamin, New York, 1962).

\bibitem{Jin} D. S. Jin and C. A. Regal, {\it Ultracold Fermi Gases}, Proceedings of the International School of Physics "Enrico Fermi", Course CLXIV, Varenna, 2006, edited by M. Inguscio, W. Ketterle, and C. Salomon (IOS Press, Amsterdam, 2008).

\bibitem{Guan} X. Guan, M. T. Batchelor, and C. Lee, Rev. Mod. Phys. {\bf 85}, 1633 (2013).

\end{thebibliography}
\end{document}